\documentclass[journal]{IEEEtran}
\ifCLASSINFOpdf
  \usepackage[pdftex]{graphicx}
  \graphicspath{{../pdf/}{../jpeg/}}
  \DeclareGraphicsExtensions{.pdf,.jpeg,.png}
\else
  \usepackage[dvips]{graphicx}
  \graphicspath{{../eps/}}
  \DeclareGraphicsExtensions{.eps}
\fi

\usepackage{graphicx}
\usepackage{graphics}
\usepackage{epsfig}\usepackage{epstopdf}
\usepackage{epstopdf}
\usepackage{stfloats}
\usepackage[cmex10]{amsmath}
\usepackage{algorithmic}
\usepackage{array}
\usepackage{mdwmath}
\usepackage{mdwtab}
\usepackage{graphicx}
\usepackage{subfigure}
\usepackage{color}
\usepackage{amsfonts,amssymb}
\usepackage[colorlinks,
            linkcolor=red, 
             anchorcolor=red, 
            citecolor=green, 
            ]{hyperref}

\usepackage{algorithm}
\usepackage{algorithmic}

\usepackage{array}
 \usepackage{mathrsfs}

\usepackage{amssymb}
\usepackage{amsmath}
\usepackage{cite}
\usepackage{url}
\usepackage{xcolor}
\usepackage{cite,graphicx,amsmath,amssymb}
\usepackage{subfigure}
\usepackage{fancyhdr}
\usepackage{mdwmath}
\usepackage{mdwtab}
\usepackage{caption}
\usepackage{amsthm}

\usepackage{mdwtab}

\usepackage{threeparttable}

\newtheorem{theorem}{Theorem}

\newtheorem{lemma}{Lemma}

\newtheorem{corollary}{Corollary}

\begin{document}
\title{Flying Car Transportation System: Advances, Techniques, and Challenges
\thanks{Manuscript received**, 2019; revised **, 2019; accepted **, 2019. The associate editor coordinating the review of this paper and approving it for publication was ***. (Corresponding author: Gaofeng Pan.)}
\thanks{G. Pan and M.-S. Alouni are with Computer, Electrical and Mathematical Sciences and Engineering Division, King Abdullah University of Science and Technology (KAUST), Thuwal 23955-6900, Saudi Arabia.}
}

\author{Gaofeng Pan, $Senior Member, IEEE$, and Mohamed-Slim Alouini, $Fellow, IEEE$ }


\maketitle

\begin{abstract}
Since the development of transport systems, humans have exploited ground-level, below-ground, and high-altitude spaces for transportation purposes. However, with the increasing burden of expanding populations and rapid urbanization in recent decades, public transportation systems and freight traffic are suffering huge pressure, plaguing local governments and straining economies. Engineers and researchers have started to re-examine, propose, and develop the underused near-ground spaces (NGS) for transportation purposes. For instance, flying cars, which are not a totally novel idea, aim at solving the traffic congestion problem and releasing the strains on existing city transport networks by utilizing unoccupied NGS. Flying cars differ from traditional grounded transportation systems that are  entirely limited by their physical space, such as trains on tracks or automobiles on roads. Flying cars do not occupy or compete for high-altitude spaces used by air traffic for long-distance transfer. However, there is a clear lack of specific literature on flying cars and flying car transportation systems (FCTS), which this paper aims to address by describing modern advances, techniques, and challenges of FCTS. We explore the inherent nature of NGS transportation and devise useful proposals to facilitate the construction and commercialization of FCTS. We begin with an introduction on the increasing need for NGS transportation and we address the advantages of using flying cars. Next, we present a brief overview of the history of development of flying cars in terms of the historic timeline and the technique development. Then, we discuss and compare the state of the art in the design of flying cars, including the take-off \& landing (TOL) modes, pilot modes, operation modes, and power types, which are related to the adaptability, flexibility \& comfort, stability \& complexity, and environment friendliness of flying cars, respectively. Additionally, since large-scale operations of flying cars can improve current transportation problems, we also introduce different facets of the various designs of FCTS, including path and trajectory planning, supporting facilities, and commercial designs. Finally, we discuss the challenges that might arise while developing and commercializing FCTS in terms of safety issues, commercial issues, and ethical issues.

\end{abstract}



\section{Introduction}
Recent decades have witnessed the rapid expansion of cities across the world, especially in developing and developed countries across Africa and Asia. Increasing levels of urbanization have resulted in significant expansion in populations and businesses, leading to high geographical functionalization of city modules. Consequently, public transportation and freight traffic have also experienced increased pressure in line with expanding economic development, and this overdevelopment has paradoxically limited further economic and social development.

\begin{figure*}[!htb]
\centering
\includegraphics[width= 7in,angle=0]{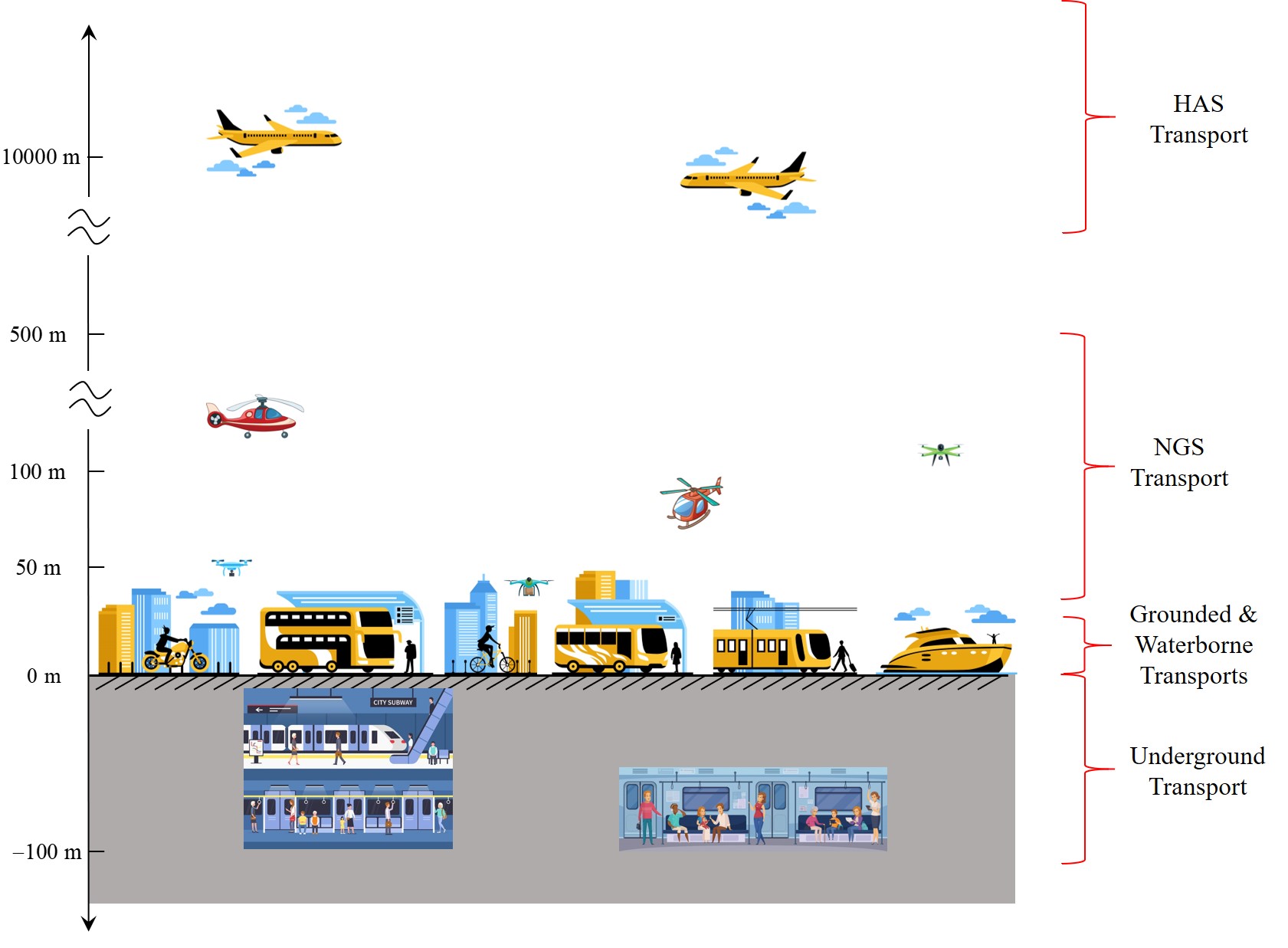}
\caption{Main transportation systems.}
\label{fig_0}
\end{figure*}

To meet transportation demands in densely urbanized environments, humans have utilized ground-level, waterborne, high-altitude spaces (HAS), and underground spaces (such as subways) for transportation systems, as depicted in Fig. \ref{fig_0}. Table \ref{CMTS} compares the main transportation systems in terms of the geographical space used. The most popular ground transportation systems, i.e., road/train transport, suffer from track and road limitations, resulting in poor flexibility and congestion, especially in urban areas. Waterborne transport, including maritime and fluvial transports, is mainly used exclusively for long-distance freight transport due to its low transport cost; however, it is not suitable for most urban environments. HAS transportation systems, i.e, air transportation, is also usually reserved for long-distance human/goods delivery, but with a significantly higher cost, and thus is also impractical for urban applications. Therefore, as indicated by Table \ref{CMTS}, only underground, ground level, and near-ground space (NGS) transport solutions are suitable for urban scenarios.

Therefore, to remove these obstacles that are preventing further development, various schemes have been proposed and implemented by cities and researchers to improve and optimize existing underground and grounded transportation systems, such as metro and public bus systems, and further to enlarge the traffic load in the face of increasing traffic demands. However, serious limitations of the traffic loads for the underground and grounded transportation systems remain. To ease the pressure on traffic loads, engineers have been exploring alternative physical spaces for the development of future transport pathways, including the unexplored NGS, ranging from tens to hundreds of meters above ground level\footnote{Until now, NGS has largely been underused except for limited helicopter travel and unmanned aerial vehicles (UAV).}. However, the use of flying cars is gathering increasing attention, and a concept that was once considered science fiction is accordingly being re-examined with serious practical proposals and developments.

\begin{table*}[!htb]

\caption{Comparisons Among The Main Transportation Systems}
\label{CMTS}
\centering
\begin{center}
\begin{tabular}{|c|c|c|c|c|c|c|}
  \hline\hline
  Types &  Distance &  Objects & Cost & Flexibility & Load& Utilization Status\\ \hline
  Underground transport& Short & People &  High & Poor&Large & Explored \\ \hline
  Grounded transport & All covered & People/Goods & Medium & Medium&Medium & Well explored\\ \hline
  Waterborne transport & Long & Goods & Low & Poor&Large & Well explored \\ \hline
  NGS transport & Short & People/Goods & Low & Good&Small & Rarely explored \\\hline
  HAS transport& Long & People/Goods & High & Good &Small& Well explored \\
  \hline\hline
\end{tabular}
\end{center}
\end{table*}

Given the lack of specific literature on flying cars and flying car transportation systems (FCTS), the main purposes of this article is to introduce, review, and discuss the past, present, and future of various aspects of FCTS. We aim to elaborate on the development history and different designs of flying cars and FCTS, and the potential challenges and prospects facing FCTS; thus, we aim to provide the community with the first comprehensive picture of FCTS in the literature.

Below we introduce the concept pf FCTS and then we discuss why FCTS is necessary, followed by a description on the organization of the content of this paper.

\subsection{What is FCTS?}
FCTS differ from traditional transportation systems by exploiting the hitherto underused NGS. FCTS are capable of expanding upon current public/private transportation by increasing the traffic load and accelerating the movement of people and goods, especially in urban areas, while simultaneously reducing traffic congestion and air pollution.

Compared with existing transportation systems, the core advantage of FCTS is the revolutionary step of systematically and commercially exploiting the NGS for the movement of people and goods, although helicopters have already utilized the NGS to some extent. Specifically, FCTS can be operated and organized to provide fast and convenient yet affordable human and freight transfer services for both public and private purposes.

\subsection{Why Should FCTS Be Introduced?}

As described above, FCTS  can be used to transfer people or goods from one location to another while operating in the NGS. Consequently, when FCTS are implemented, some essential infrastructure for traditional ground-based transportation systems, such as roads and tunnels, can be entirely avoided. Furthermore, electric-powered rotors of flying cars can be designed and equipped to dramatically reduce the on-board noise and emissions, and thus make the travel experience more comfortable for passengers. Finally, flexible operating mechanisms can be implemented and launched in FCTS, thanks to the freedom arising from the inherent system characteristics of FCTS.

Below we outline some of the unique properties of FCTS:

1) Environmentally friendly transportation is supported, as flying cars equipped with power cells with zero direct discharges and emissions can be theoretically achieved;

2) Congestion free transportation is possible by benefiting from the underexploited and unlimited NGS resource;

3) Flexible and fast door-to-door transportation is achievable because shorter travel paths can be realized by flying cars that are not obstructed by physical infrastructure, unlike ground-based transportation systems;

4) Less ground supporting infrastructure is required; most FCTS are operated in the near-ground airspace and the vertical take-off \& landing (VTOL) mode is adopted as the main and popular mode for flying cars;

5) Less competition for space among road users frees up more ground-level space for people, making cities more comfortable for pedestrians and citizens;

6) Less construction and upkeep costs are needed to operate FCTS because no roads or tunnels are required to be constructed and maintained, leading to low-cost commercial operations for FCTS.

Given these advantages, we conclude that FCTS present a promising solution to relieve traffic congestion in cities and to promote urbanization at a lower cost.

\subsection{The Organization of This Article}
The rest of this paper is organized as follows. In Section II, we present a brief review of the history of the development of flying cars. In Section III and IV, we elaborate on the designs of flying cars and FCTS, respectively. The challenges and trends of FCTS are discussed and summarized in Section V. The paper is concluded in Section VI.

\section{The History of Development of Flying Cars}
The concept of flying cars is not a novel idea and can be traced back to the beginning of the last century. In this section, we present a brief introduction to the history of development of flying cars in terms of the timeline of development and technique categories.

In the early 1900s, the airplane was first successfully invented and manufactured, while at the same time, the automobile industry was undergoing mass expansion and production through standard mass-production techniques. Many inventors attempted to combine the airplane and motorcar, as highlighted by Henry Ford in 1940: ``Mark my word: a combination of airplane and motorcar is coming. You may smile, but it will come¡± \cite{wall}. Numerous endeavors were carried out to find  suitable and practical engineering proposals; for example, Glen Curtiss' Aeroplane debuted in 1917 \cite{Curtiss}, the Waterman Aerobile in 1937 \cite{Meaden}, the ConVairCar Model 118 in 1947 \cite{NYT}, the Aero-Car series from 1949 to 1977 \cite{Gilmore,AerocarIII}, and the AVE Mizar in 1973 \cite{Van}.

\begin{figure}[!t]
\centering
\includegraphics[width= 2.8in,angle=0]{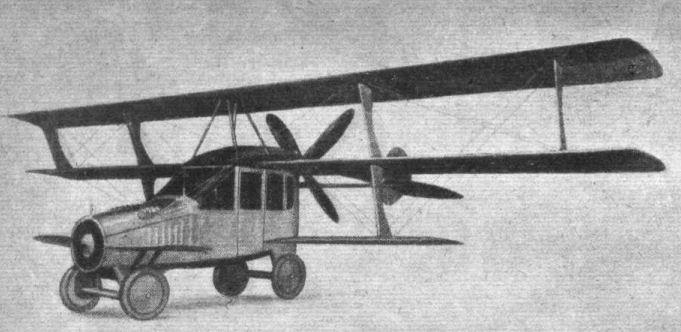}
\caption{Curtiss' Aeroplane in 1917. (\url{https://upload.wikimedia.org/wikipedia/commons/b/b6/Curtiss_Autoplane_1917.jpg.})}
\label{fig_1}
\end{figure}

\begin{figure}[!t]
\centering
\includegraphics[width= 2.8in,angle=0]{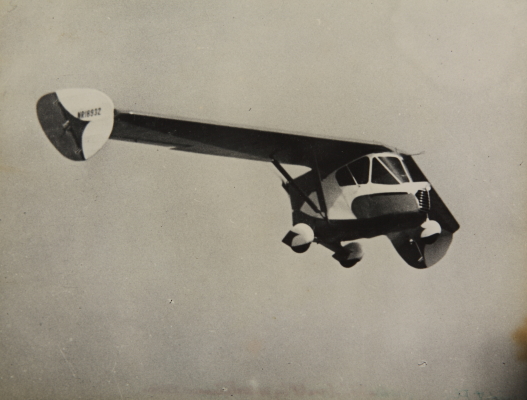}
\caption{The Waterman Aerobile in 1937. (\url{https://en.wikipedia.org/wiki/Waterman_Arrow-bile\#/media/File:Waterman_Aerobile_in_flight.jpg}.)}
\label{fig_2}
\end{figure}

\begin{figure}[!t]
\centering
\includegraphics[width= 2.8in,angle=0]{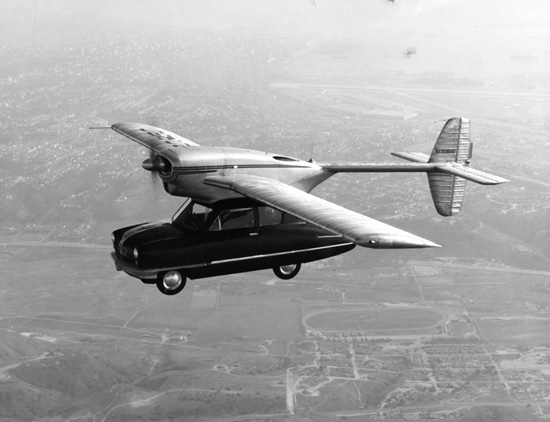}
\caption{The ConVairCar Model 118 in 1947. (\url{https://upload.wikimedia.org/wikipedia/commons/thumb/b/b5/ConvairCar_Model_118.jpg/450px-ConvairCar_Model_118.jpg})}
\label{fig_3}
\end{figure}

\begin{figure}[!t]
\centering
\includegraphics[width= 2.8in,angle=0]{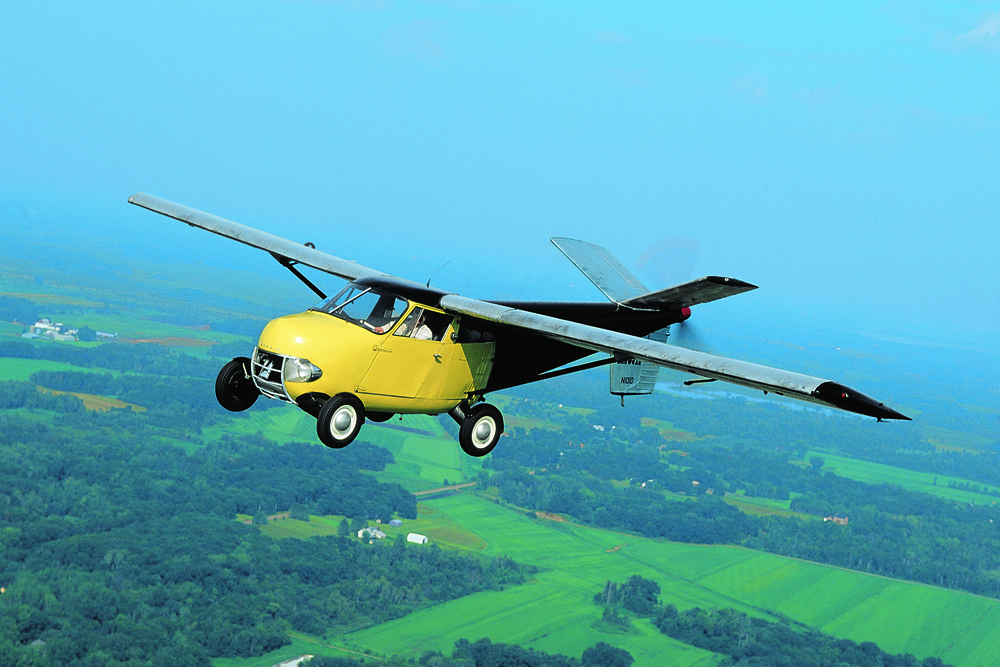}
\caption{Aero-Car Model I in 1954. (\url{https://barrettjacksoncdn.azureedge.net/staging/carlist/items/Fullsize/Cars/236076/236076_Front_3-4_Web.jpg})}
\label{fig_4}
\end{figure}

\begin{figure}[!t]
\centering
\includegraphics[width= 2.8in,angle=0]{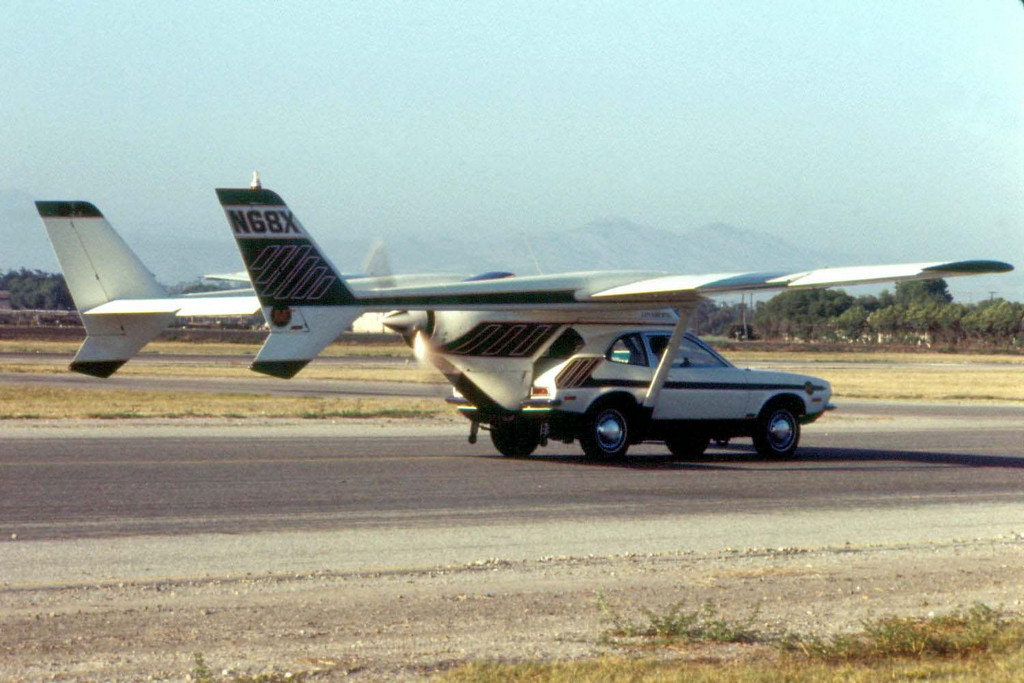}
\caption{The AVE Mizar in 1973. (\url{https://upload.wikimedia.org/wikipedia/commons/thumb/8/80/AVE-Mizar-1973-N68X-XL.jpg/450px-AVE-Mizar-1973-N68X-XL.jpg})}
\label{fig_5}
\end{figure}

\begin{figure}[!t]
\centering
\includegraphics[width= 2.8in,angle=0]{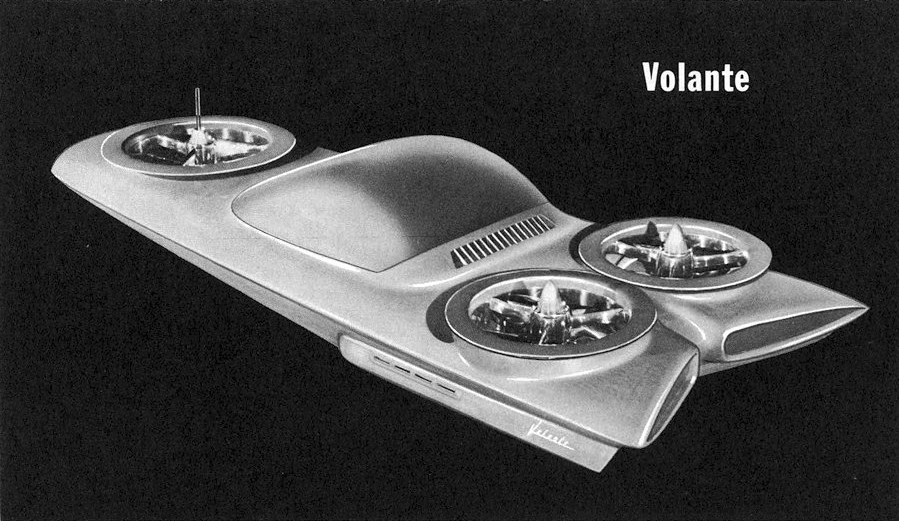}
\caption{The Volante Tri-Athodyne in 1957. (\url{http://www.carstyling.ru/resources/concept/1957-Ford-Volante-Concept-Car-Model-01.jpg})}
\label{fig_6}
\end{figure}

\begin{figure}[!t]
\centering
\includegraphics[width= 2.8in,angle=0]{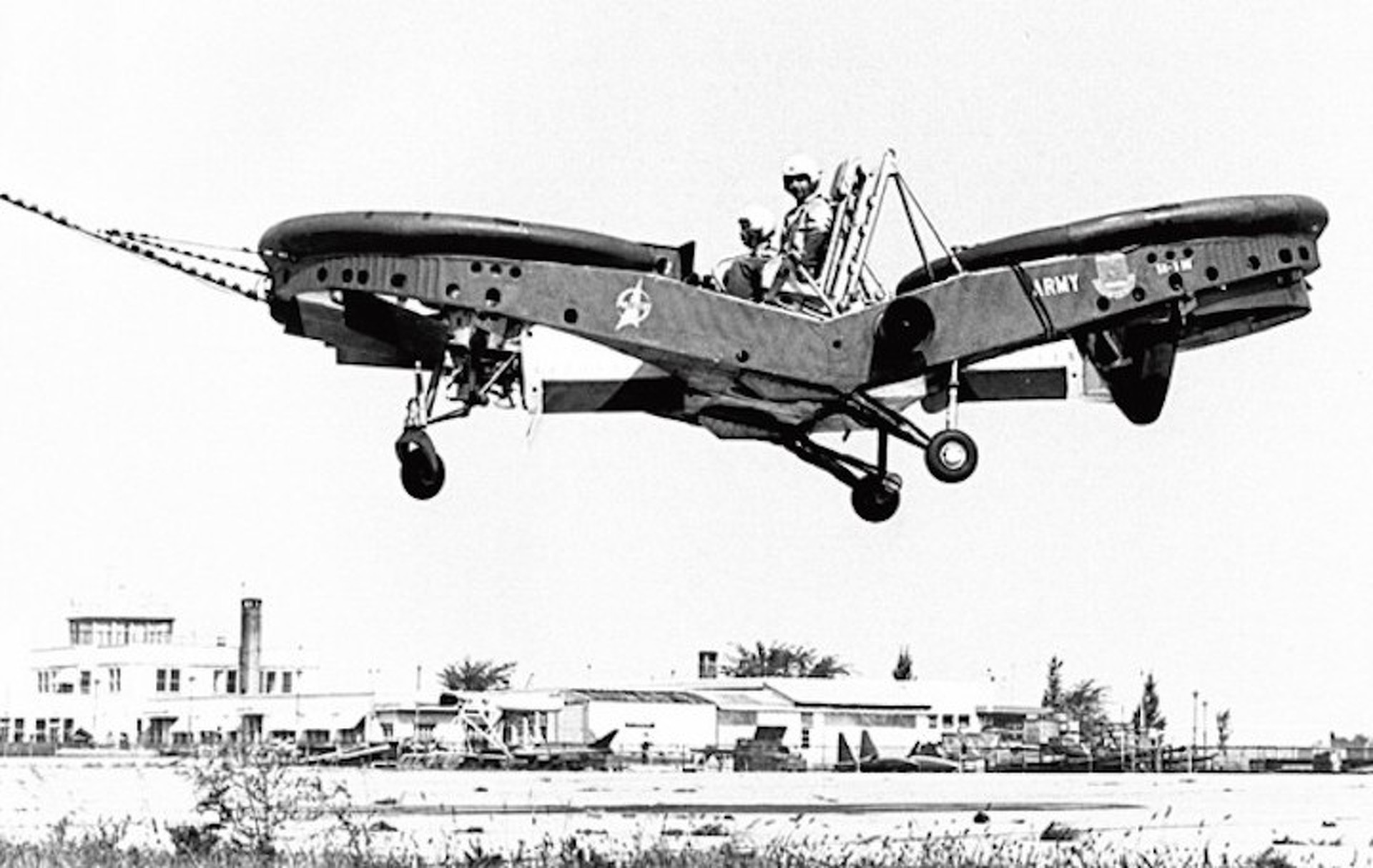}
\caption{Piasecki's VZ-8 Airgeep in 1962. (\url{https://cdn.motor1.com/images/mgl/XJP9E/s1/the-armys-quest-for-a-flying-jeep.jpg})}
\label{fig_7}
\end{figure}

\begin{figure}[!t]
\centering
\includegraphics[width= 2.8in,angle=0]{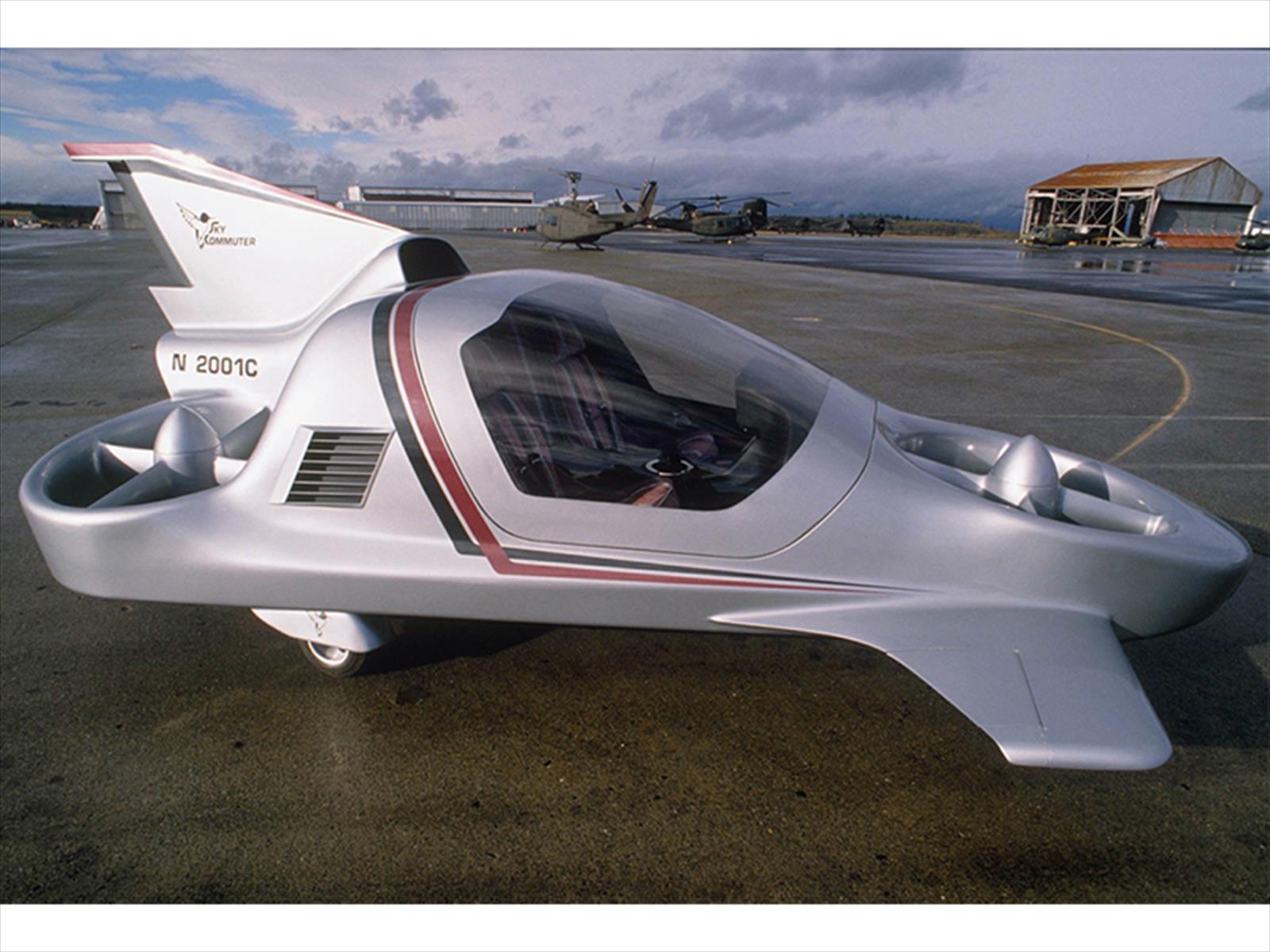}
\caption{The Sky Commuter in the 1980¡¯s. (\url{https://www.arrajol.com/sites/default/files/2015/07/27/3-GAZ_da6aa2ce0ca341379d5f91a18013e38d.jpg})}
\label{fig_8}
\end{figure}

The prototypes of flying cars depicted in Figs. \ref{fig_1}-\ref{fig_5} essentially are all equipped with fixed wings that are attached to automobiles that enable them to take-off, fly, and land.

At the same during the last century, another technical branch used duct fans to achieve vertical take-off \& landing (VTOL), such as the Volante Tri-Athodyne in 1956-1958 designed by Ford \cite{Corn}, Piasecki's VZ-8 Airgeep in 1958-1962 by Piasecki Aircraft \cite{Piasecki}, and the Sky Commuter in the 1980s by Fred Barker \cite{Zoltan}.

Unfortunately, all these attempts to establish flying cars failed, not only because of the immaturity of the related technologies at the time but also because the relatively advanced and expensive designs were beyond the practical needs and affordability of the potential consumers.

However, this century has witnessed rapid advancements in power battery technology, metallic and non-metallic materials, autonomous control, and mechanical manufacturing, and thus the development of flying cars has advanced significantly in recent years. Also, the commercialization of flying cars has been motivated by increased transportation pressure, rapid urbanization, and global economic growth. Therefore, to date, the current technical trend of developing flying cars is driven by a need to provide safe, green, fast, and convenient human/freight transportation services in urban areas by using electric power, VTOL operation mode, and autonomous piloting techniques.

Several new start-up technology companies, such as ride-hailing companies (e.g., Uber), and some famous automobile and aircraft manufacturing companies, such as Rolls-Royce, Audi, Geely, Tesla, Toyota, Hyundai, Airbus, and Boeing, are currently engaged in research into flying car technology, and various types of flying cars are being designed, manufactured, and tested \cite{palv}-\cite{Audi}, which we will discuss in detail in the following sections\footnote{We do not present photos of the main modern designs of flying cars due to page limitations, but we provide links \cite{palv}-\cite{Audi} to some manufacturers that are leading the way in the commercialization of flying cars. For example, Uber is working with its Elevate Network partners to launch fleets of small, electric VTOL aircraft in Dallas, Los Angeles, and Melbourne \cite{uber}, and Morgan Stanley research says that ``accelerating tech advances and investment could create a \$1.5 trillion market by 2040" \cite{Morgan}.}.

\section{Flying Car Design}
\begin{figure}[!t]
\centering
\includegraphics[width= 3.7in,angle=0]{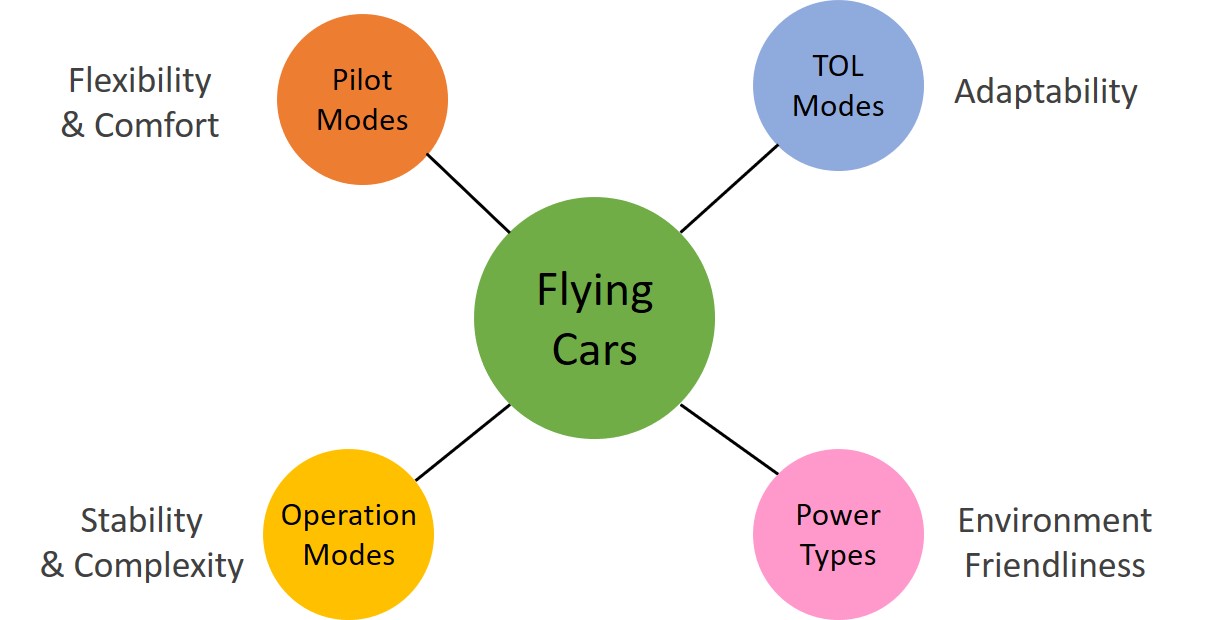}
\caption{The design aspects of flying cars. }
\label{fig_9}
\end{figure}

In this section, we discuss the design of flying cars in terms of the following four aspects: Take-off \& landing (TOL) modes, pilot modes, operation modes, and power types, which are  related to the adaptability, flexibility \& comfort, stability \& complexity, and environmental friendliness of flying cars, respectively (as presented in Fig. \ref{fig_9}).

\subsection{TOL Modes} Generally, there are four kinds of TOL modes for aerial vehicles:

\subsubsection{VTOL (Vertical Take-Off Landing) Mode} Aerial vehicles can take off and land vertically. Under this mode, no runways are needed for either take-off or landing, thus giving more choice for potential TOL sites and more freedom to reach passengers anywhere that they wish to go.

\subsubsection{VTHL (Vertical Take-Off Horizontal Landing) Mode} Aerial vehicles can take off vertically but return to the ground by taxiing along a runway. Thus, runways are necessary for the landing process, but take-off is not restricted to a runway.

\subsubsection{HTVL (Horizontal Take-Off Vertical Landing) Mode} Aerial vehicles taxi along the runway before take-off, and land vertically at the destination. Thus, runways are required for the take-off process, but there is flexibility for the landing process.

\subsubsection{HTOL (Horizontal Take-Off Landing) Mode} Aerial vehicles require a runway for both take-off and landing. This mode is the most common mode for fixed-wing aircraft, and runways or ground supports are always required.

Table \ref{TOL} presents the comparisons between these four TOL modes. The VTOL mode is evidently the best choice for flying cars that operate in urban districts, where the space to build runways is not normally available. The HTOL mode is a good option for rural areas, where there is sufficient space to build runways.

\begin{table*}[!htb]

\caption{Comparisons Among The Four TOL Modes}
\label{TOL}
\centering
\begin{center}
\begin{tabular}{|c|c|c|c|c|c|}
  \hline\hline
  Types & Technical Complexity & Runways/Supports & Maintenance Cost & Rotary Wings/Vertical Fans & Fixed Wing\\ \hline
  VTOL & High & Not Required &  High & Yes & No \\ \hline
  VTHL & Medium & Required & Medium & Yes & Yes\\ \hline
  HTVL & Medium & Required & Medium & Yes & Yes \\ \hline
  HTOL & Low & Required & Low & No & No \\
  \hline\hline
\end{tabular}
\end{center}
\end{table*}
\begin{table*}[!htb]
\caption{TOL Modes of The Main Modern Flying Cars}
\label{FC}
\centering
\begin{center}
\begin{tabular}{|c|c|c|c|c|c|}
  \hline\hline
  Flying Car & Company & TOL & Rotary Wings/Fans & Propeller & Fixed/Foldable Wing \\\hline
    PAL-V Liberty \cite{palv}& PAL-V & HTOL & Yes & Yes& No\\ \hline
  Elevate \cite{uber}& Uber & VTOL & Yes &  No & No \\ \hline
  Heaviside \cite{kittyhawk}&Kitty Hawk& HTOL & No & Yes & Yes \\ \hline
  Flyer \cite{kittyhawk} &Kitty Hawk &VTOL & Yes & No & No \\ \hline
  The CityAirbus \cite{RR}& Rolls-Royce \& Airbus & VTOL & Yes & No & No \\ \hline
  Vahana \cite{airbus}& Airbus& VTOL & Yes & No & No \\ \hline
  AeroMobil V. 4.0 \cite{aeromobil} & AeroMobil& HTOL& No & No & Yes \\ \hline
  AeroMobil V. 5.0 \cite{aeromobil}& AeroMobil& VTOL& Yes & No & No \\ \hline
  VoloCity \cite{volocopter}& Volocopter& VTOL& Yes & No & No \\ \hline
  MOOG \cite{SureFly}& SureFly& VTOL& Yes & No & No \\ \hline
  BlackFly \cite{Opener}& Opener& VTOL& Yes & No & No\\ \hline
  The Transition \cite{terrafugia}& Terrafugia& HTOL& No & Yes & Yes \\ \hline
  The TF-2 \cite{terrafugia}& Terrafugia& VTOL& Yes & No & No \\ \hline
  EHang AAV \cite{ehang}& EHang& VTOL& Yes & No & No \\ \hline
  Cora  \cite{wisk}& Wisk&VTOL&Yes & Yes & Yes \\ \hline
  Joy Aviation Air Taxi \cite{jobyaviation}& Joby Aviation& VTOL& Yes & No & No \\ \hline
  Jaunt Aircraft \cite{jauntair}& Jaunt Air Mobility & VTOL& Yes & Yes & No \\ \hline
  Passenger Air Vehicle\cite{boeing}& Boeing NeXt & VTOL& Yes & Yes & Yes \\ \hline
  Cargo Air Vehicle\cite{boeing}& Boeing NeXt & VTOL& Yes & No & No \\ \hline
  SD-XX\cite{skydrive}& SkyDrive &VTOL& Yes & No& No\\ \hline
  The Volante Vision Concept\cite{astonmartin}& Aston Martin & VTOL& Yes & No & No \\ \hline
  Moller M400 Skycar\cite{moller}& Moller International & VTOL& Yes & No & No \\\hline
  The Nexus\cite{Bell}& Bell&HTOL& Yes & No & No \\\hline
  WD-1\cite{ detroit}& Detroit Flying Cars &VTOL & Yes & Yes & Yes \\\hline
  DR-7\cite{Delorean}& Delorean Aerospace &VTOL & Yes& No & Yes \\\hline
  The Pop.Up Next\cite{Audi}& Audi &VTOL & Yes& No & Yes \\\hline
    \hline
\end{tabular}
\end{center}
\end{table*}

\begin{table*}[!htb]
\caption{Comparisons Among The Main Modern Flying Cars}
\label{FM}
\centering
\begin{center}
\begin{tabular}{|c|c|c|c|c|c|c|}
  \hline\hline
  Flying Car & Speed (km/h)&Range (km)& Operation Modes & Power Types & Control Modes &Passengers\\\hline
  PAL-V Liberty \cite{palv}& 160$^{{\rm{T}}0}$& 400$^{{\rm{T}}1}$ & Helicopter-car & Hydrocarbon & HP& Multiple\\ \hline
  Elevate \cite{uber}& 241.4$\sim$370.15$^{{\rm{T}}2}$&321.87$^{{\rm{T}}3}$ & Helicopter-airplane & Electric &  HP & Multiple\\ \hline
  Heaviside \cite{kittyhawk}&------&------& Airplane & Electric & HP & Single\\ \hline
  Flyer \cite{kittyhawk} &------&------ &Helicopter & Electric & HP & Single\\ \hline
  The CityAirbus \cite{RR}&120.70$^{{\rm{T}}0}$ &------ & Helicopter & Electric & SP & Multiple\\ \hline
  Vahana \cite{airbus}&190 &50 & Helicopter-airplane & Electric & SP & Single\\ \hline
  AeroMobil V. 4.0 \cite{aeromobil} &360$^{{\rm{T}}0}$ & 750& Airplane-car& Hybrid & HP& Multiple\\ \hline
  AeroMobil V. 5.0 \cite{aeromobil}& ------&------& Helicopter-car& Electric & Hybrid& Multiple\\ \hline
  VoloCity \cite{volocopter}&80$\sim$100$^{{\rm{T}}4}$&30-35& Helicopter& Electric & Hybrid& Multiple\\ \hline
  MOOG \cite{SureFly}& 120.7 & ------& Helicopter& Electric &Hybrid& Multiple\\ \hline
  BlackFly\cite{Opener}& $>$128.75$^{{\rm{T}}5}$ & 64.37$^{{\rm{T}}6}$& Helicopter-airplane& Electric  & HP & Single\\ \hline
  The Transition \cite{terrafugia}& 161$^{{\rm{T}}0}$ & 644 & Airplane-car& Hydrocarbon$^{{\rm{T}}7}$ & HP& Multiple\\ \hline
  The TF-2 \cite{terrafugia}& 230$^{{\rm{T}}0}$& 300& Helicopter-airplane& Hydrocarbon$^{{\rm{T}}8}$& HP& Multiple\\ \hline
  EHang AAV \cite{ehang}& 130$^{{\rm{T}}9}$& ------& Helicopter& Electric & SP& Multiple\\ \hline
  Cora  \cite{wisk}& 160 &40.23& Helicopter-airplane& Electric & SP& Multiple\\ \hline
  Joy Aviation Air Taxi \cite{jobyaviation}& 321.87 & 241.4&Helicopter-airplane & Electric & HP& Multiple\\ \hline
  Jaunt Aircraft \cite{jauntair}& 281.64$^{{\rm{T}}9}$ &------& Helicopter-airplane & Electric & HP& Multiple\\ \hline
  Passenger Air Vehicle\cite{boeing}& ------&------ &Helicopter-airplane & Electric& Hybrid& Multiple \\ \hline
  Cargo Air Vehicle\cite{boeing}& ------&------ &Helicopter& Electric & SP& None\\ \hline
  SD-XX\cite{skydrive}& 60 & 20-30 & Helicopter& Electric & Hybrid& Multiple\\ \hline
  The Volante Vision Concept\cite{astonmartin}& ------&------ & Helicopter& Hybrid & SP& Multiple\\ \hline
  Moller M400 Skycar\cite{moller}& 533 &1213 & Helicopter& Electric & HP & Multiple \\\hline
  The Nexus\cite{Bell}& 241.4$^{{\rm{T}}0}$ &241.4 & Airplane& Hybrid & HP & Multiple \\\hline
  WD-1\cite{detroit}& 201.17 &643.74 & Airplane-car& Hybrid & Hybrid & Multiple \\\hline
  DR-7\cite{Delorean}& 241 &193 & Helicopter& Electric & SP & Multiple \\\hline
  The Pop.Up Next\cite{Audi}& ------ &------ & Helicopter-car& Electric & SP & Multiple \\\hline
  \hline
\end{tabular}
\begin{tablenotes}
    \item T0: The maximum speed;
    \item T1: The maximum range is 400 km and 500 km for two and a single person, respectively;
    \item T2: A desirable speed for VTOL vehicle;
    \item T3: Obtained at the best range flight speed;
    \item T4: The optima cruise speed range with the trade-off between time efficiency, and cost, safe operation at low altitudes and low noise;
    \item T5: The speed is no more than 99.78 km/h for the USA;
    \item T6: Obtained with the load 200 lbs, while the rang is 40.23 km for the USA with load 200 lbs;
    \item T7: Hybrid-electric motors for drive mode;
    \item T8: Under air flying mode, 8 electric motors are powered initially by a turbine generator and ultimately powered by batteries, while for drive mode it will be hybrid-electrically powered initially, and become purely electric as practical;
    \item T9: The maximum cruise speed.
   \end{tablenotes}
\end{center}
\end{table*}

Flying cars should be equipped with rotary wings or vertical fans to achieve VTOL, while fixed/fold-able wings are required for HTOL. From a technical viewpoint, rotary wings or vertical fans are more complex and expensive than fixed/foldable wings. The best choice of TOL mode for a flying car depends on the application scenario.

Table \ref{FC} compares the TOL modes employed for the main types of flying cars that are currently under development. VTOL is the most popular, with 21 prototypes adopting VTOL. Only five prototypes choose HTOL because most flying cars in development are aimed at the transportation market in urban areas where there is limited space for TOL operation. Moreover, as shown in Table \ref{FC}, rotary wings are required to realize VTOL, which, because of their technical complexity, unavoidably increases the costs of VTOL compared to HTOL. Consequently, a commercially viable solution is required to promote flying cares with VTOL for industrialization and commercialization.

On the contrary, as mentioned before, flying cars with HTOL are a suitable commercial choice for rural areas where there is sufficient space to build runways for taxiing.

\subsection{Pilot Modes}
\begin{figure}[!t]
\centering
\includegraphics[width= 1.7in,angle=0]{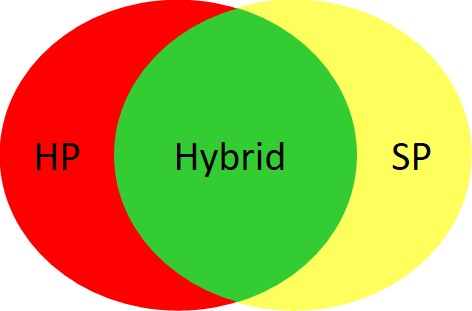}
\caption{Pilot modes of flying cars. }
\label{fig_10}
\end{figure}

Pilot modes refer to how the flying cars are operated and controlled during TOL and flying, which is most important in terms of the hardware costs, operating costs (e.g., fuel consumption and piloting costs), and the safety of flying cars. Typically, there are three kinds of control modes for flying cars: Human-piloted (HP), self-piloted (SP), and hybrid modes, the relationship of which is depicted in Fig. \ref{fig_10}. We discuss the three kinds of control modes below.
\subsubsection{HP Mode}
Under this mode, flying cars are operated by pilots, as in traditional air-crafts, and a driver's license or a pilot certificate is required. Thus, labor costs are significant (through training fees and wages paid to the pilots, for example), leading to a considerable burden on the commercial operation of flying cars. The ratio of driver/pilot to passengers per flying car is much lower compared to traditional public air-crafts. However, the human-pilot mode offers high operational freedom and the flexibly to react to some emergency situations.

\subsubsection{SP Mode}
Under the SP mode, flying cars can perform all operations in a fully autonomous way without the need for pilots, thus incurring lower operating costs than the HP mode. Currently, the SP mode is the most popular choice for flying cars, as shown in Table \ref{FM}, where the SP mode is supported by 14 of the 26 kinds of flying cars currently under development (including those with the hybrid mode). The HP mode benefits from the maturity of the technology surrounding autonomous control and navigation that is making SP vehicles smarter and safer. Moreover, in the future, artificial intelligence technologies will be more reliable and efficient. For the commercial application of flying cars, the SP mode is especially suitable for public transport air shuttles with fixed routes.

\subsubsection{Hybrid Mode}
The hybrid mode has been proposed and adopted by some flying cars under development to take full advantage of the individual benefits of two control modes mentioned above. Although the manufacturing costs of flying cars with the hybrid mode are higher than the costs of the two other modes, the hybrid mode is still the best candidate for certain applications such as flexible transportation without fixed routes. For example, flying cars using the hybrid mode are well suited for tours and sightseeing spots.

\subsection{Operating Modes}
Because various kinds of TOL modes and types of wings exist, flying cars can be endowed with different operating modes, such as airplane, helicopter, and hybrid modes.
\subsubsection{Airplane mode}
Only two flying car prototype in development operate like traditional airplanes, i.e., the Heaviside produced by Kitty Hawk \cite{kittyhawk} and the Nexus by Bell \cite{Bell}. For example, the Heaviside can take-off, fly, and land like traditional airplanes, but can not travel on roads like cars. Therefore, considerable space is required for the TOL of this kind of flying car, making it unsuitable for urban areas.

\subsubsection{Helicopter Mode}
The helicopter mode is a popular operation mode for flying cars, which fully inherits the TOL and flying characteristics of helicopters. Thanks to the unique TOL procedure of the helicopter mode that does not require a runway and thus can save a lot of space, flying cars with this mode are quite suitable for transportation in urban areas. As suggested in Table \ref{FM}, there are 10 types of flying cars that adopt the helicopter mode, i.e., Flyer by Kitty Hawk \cite{kittyhawk}, the CityAirbus by Rolls-Royce \& Airbus \cite{RR}, Volocity by Volocopter \cite{volocopter}, MOOG by SureFly \cite{SureFly}, EHang AAV by EHang \cite{ehang}, Cargo Air Vehicle by Boeing NeXt \cite{boeing}, SD-XX by SkyDrive \cite{skydrive}, the Volante Vision Concept by Aston Martin \cite{astonmartin}, Moller M400 Skycar by Moller International \cite{moller}, and DR-7 by Delorean Aerospace \cite{Delorean}.

\subsubsection{Hybrid Mode}
To satisfy different application demands, designers have also presented some other hybrid modes that combine the advantages of individual operating modes.

a) Helicopter-car mode: Under this mode, flying cars can take-off, fly, and land like helicopters, and also travel on roads like traditional cars. Generally, this kind of flying car is equipped with foldable rotary wings. The helicopter-car mode is applied on PAL-V Liberty by PAL-V \cite{palv}, AeroMobil V. 5.0 by AeroMobil \cite{aeromobil}, and the Pop.Up Next by Audi \cite{Audi}.

b) Helicopter-airplane mode: Under this mode, flying cars take-off and land like helicopters and fly like airplanes. The main difference between the helicopter-airplane mode and the helicopter mode is that fixed/foldable wings and horizontal propellers are employed to improve the flying performance. As the helicopter-airplane mode shows high adaptability for urban areas, it is the most popular operation mode, with eight flying cars adopting this mode currently being proposed/designed by manufacturers, as shown in Table \ref{FM}.

c) Airplane-car mode: Foldable wings are employed by this kind of flying cars to improve both flying performance in the air and running performance on the road. The flying cars can taxi on regular roads for take-off and landing. Because of the space requirements for this TOL mode, the flying cars under airplane-car mode are not usually suitable for urban areas, where normally there is insufficient space to build runways; thus, this mode is probably a better choice for rural areas or large outdoor scenic sites. As shown in Table \ref{FM}, only three types of flying cars have adopted the airplane-car mode: AeroMobil V. 4.0 by AeroMobil \cite{aeromobil}, the Transition by Terrafugia \cite{terrafugia}, and WD-1 by Detroit Flying Cars \cite{ detroit}.

\subsection{Power Types}
Depending on the application purposes, there are three main power types for flying cars: electric power, hydrocarbon fuel, and hybrid power.

\subsubsection{Electric Power}
Electric propulsion is widely the preferable propulsion method for flying cars because it has many desirable characteristics such as lower emissions and less noise pollution. Electric batteries are the preferable energy source for flying cars in urban areas (19 out of 26 kinds of flying car designs use electric power presented in Table \ref{FM}). There are some specific requirements for batteries, such as the energy density (the amount of energy per unit weight), the recharging time (time needed to recharge), the cycle life (the number of complete discharge/charge cycles the battery can support before the capacity falls below 80\% of its original capacity), and the average price of the battery (the cost per kilowatt-hour), which are important to support frequent use for urban transport. Finally, the pollution that is produced from the heavy metals and acids of wasted batteries should be considered. Thus, green battery technologies with high energy density, large cycle life, and low cost are necessary when designing flying cars.

\subsubsection{ Hydrocarbon Fuel}
Some flying car designs use hydrocarbon fuel and combine the advantages of traditional cars and aircraft by enabling flying cars to travel on the ground, like traditional automobiles, and to fly in the air like airplanes/helicopters; see, for example, PAL-V Liberty by PAL-V \cite{palv}, and the Transition and TF-2 by Terrafugia \cite{terrafugia}. Flying cars that use hydrocarbon fuel carry the same disadvantages as regular vehicles and aircraft in terms of noise and air pollution produced by the internal-combustion engine, as well as climate change considerations. However, the hardware costs and manufacturing complexity of flying cars that are powered by hydrocarbon fuel are substantially lower than flying cars that use hybrid power, as will be discussed below.
\begin{figure}[!t]
\centering
\includegraphics[width= 3.5in,angle=0]{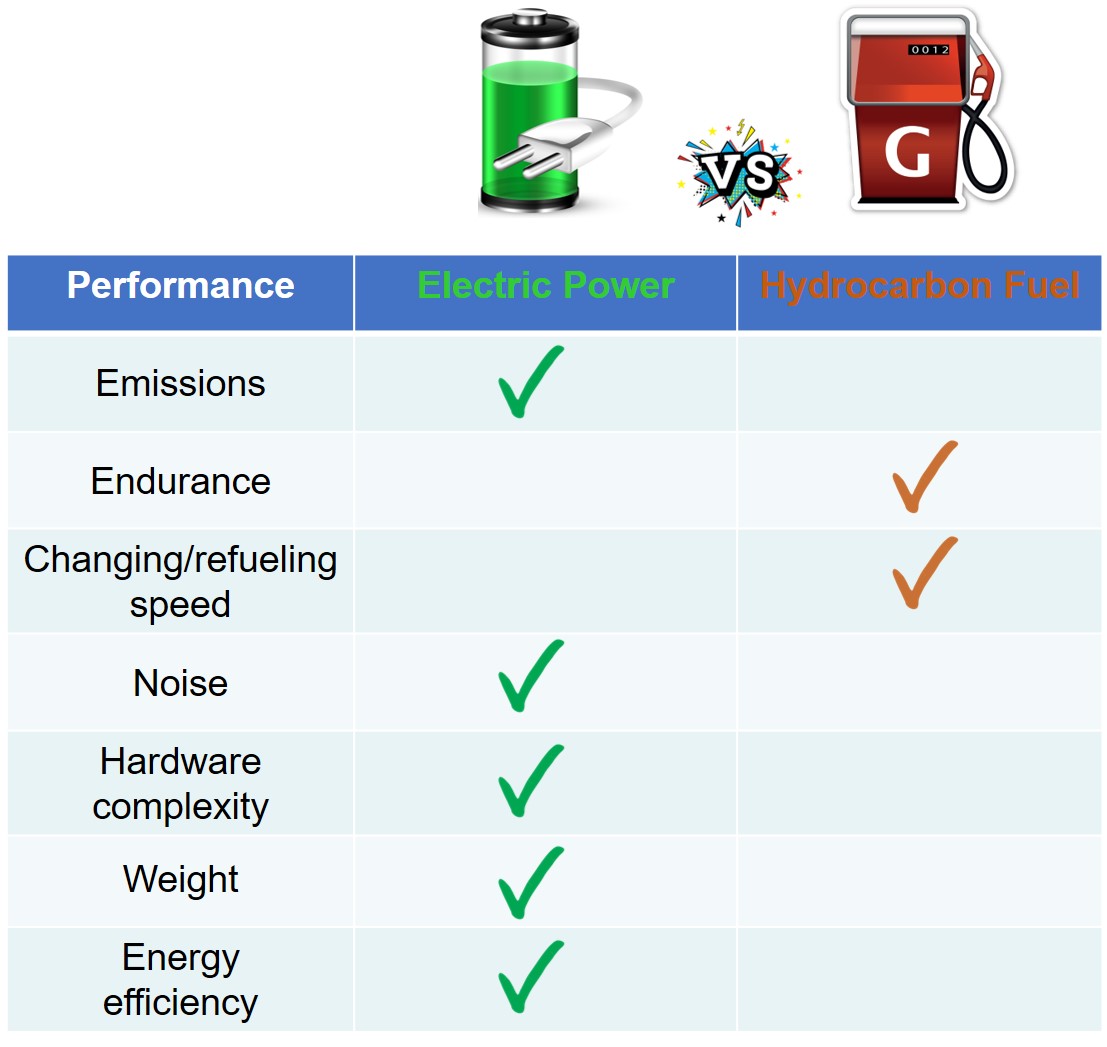}
\caption{Comparison of electric power and hydrocarbon fuel. }
\label{fig_11}
\end{figure}
\subsubsection{Hybrid Power}
As shown in Fig. \ref{fig_11}, electric power outperforms hydrocarbon fuel in terms of reduced emissions, noise, hardware complexity, weight, and increased energy efficiency, confirming that electric power is more suitable for flying cars, especially in urban areas. However, the durability and refueling speeds of hydrocarbon fuel are superior to electric power, making hydrocarbon fuel more suitable for long-distance transportation designs. The problem of the slow charging speed of electric power can be addressed by replacing the batteries at TOL sites rather than waiting for the depleted battery to recharge.

A hybrid power option has been proposed for flying cars, similar to that employed in the automobile industry, to combine the advantages of both electric power and hydrocarbon fuel. However, such a design will inevitably increase the hardware complexity and the cost of flying cars, resulting in poor economic competitiveness.

Table \ref{FM} lists some existing designs of flying cars, showing that hybrid power is the best power configuration for hybrid transportation on the ground in cars, which has been utilized by AeroMobil V. 4.0 by AeroMobil \cite{aeromobil}, the Volante Vision Concept \cite{astonmartin}, the Nexus by Bell \cite{Bell}, and WD-1 by Detroit Flying Cars \cite{ detroit}. For example, AeroMobil V. 4.0 by AeroMobil uses the hybrid power configuration on the ground, which is powered by a hybrid-electric system, and the same engine has been employed to power the vehicle in the air under flight mode, which in turn powers a pair of electric motors located in the front axle \cite{aeromobil}.

Thus, we can conclude that, while electric power is the greenest and most cost-saving power configuration for flying cars, hybrid power is the more reasonable commercial choice for hybrid transportation, especially over long distances.

\section{The Design of FCTS}
In the previous section, we discussed the issues related to the design of flying cars. In this section, we  introduce FCTS in terms of the following three aspects: Path and trajectory planning, the required supporting facilities, and the commercial designs, which are the key problems encountered in the large-scale operation of FCTS.

\subsection{Path/Trajectory Planning}
\begin{figure}[!t]
\centering
\includegraphics[width= 3.5in,angle=0]{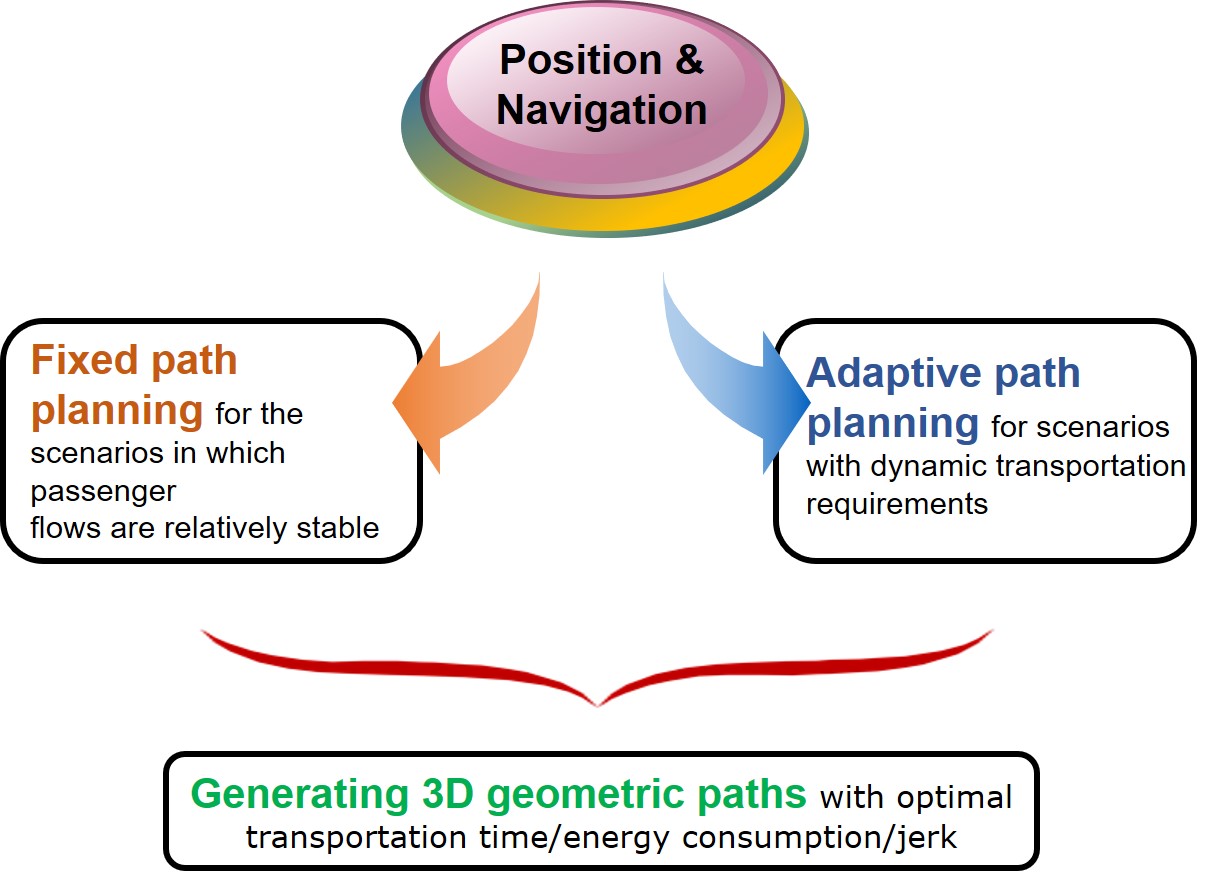}
\caption{Path planning for FCTS. }
\label{fig_12}
\end{figure}

Path/trajectory planning is a crucial issue in the field of FCTS. Indeed, flying cars are mainly designed to operate in urban areas to ultimately achieve shorter transportation times and relieve road congestion. However, high operating speeds in the air may cause collisions and consequent chaos among the flying cars and buildings; thus, extremely accurate scientifically guided path and trajectory planning is essential. However, to date, there are no provisions for FCTS route planning yet, especially for dedicated VTOL paths/routes. Therefore, particular care should be given to generating flying paths/trajectories that can be executed in the air above a city, while being simultaneously efficient and safe for transportation, in terms of avoiding any unnecessary flying and possible security vulnerabilities. The content architecture of this subsection is given in Fig. \ref{fig_12}.
\subsubsection{Position \& Navigation}
During the flight process, the most important thing for a flying car following a predesigned path is  accurate positioning and navigation. Therefore, achieving precise information on position and navigation is essential.

Various methods have been developed to obtain two-dimensional positional information of an object, such as global positioning system (GPS) and cellular network/wireless fidelity (Wi-Fi). These technologies have been applied to numerous ground movement scenarios. However, because flying cars normally operate at heights ranging from tens to hundreds of meters, accurate three-dimensional positioning information is required for FCTS to precisely determine where a flying car is and at which height the flying car is operating at.

Radars are an effective tool for ground control centers. Accurate positioning can be achieved by using new position and navigation algorithms that adopt some reference nodes, such as base stations on the roof of buildings/hills and low-attitude platforms, for example, which serve as aerial access points for 5th or future generation wireless communications.

\subsubsection{The Objects of Path/Trajectory Planning}
The purpose of path/trajectory planning for FCTS is to generate 3D geometric paths, from an initial to a final destination point, passing through predefined through-points, either in restricted spaces identified by governments or in wildlife conservation areas.

The algorithms for path/trajectory planning are usually categorized according to the functions that are optimized in terms of transportation time/energy consumption/jerk (i.e., the derivative of the acceleration, which is related to the passenger experience), which are devoted to satisfying the requirements in the application scenarios of flying cars.
\begin{figure*}[!htb]
\centering
\includegraphics[width= 5in,angle=0]{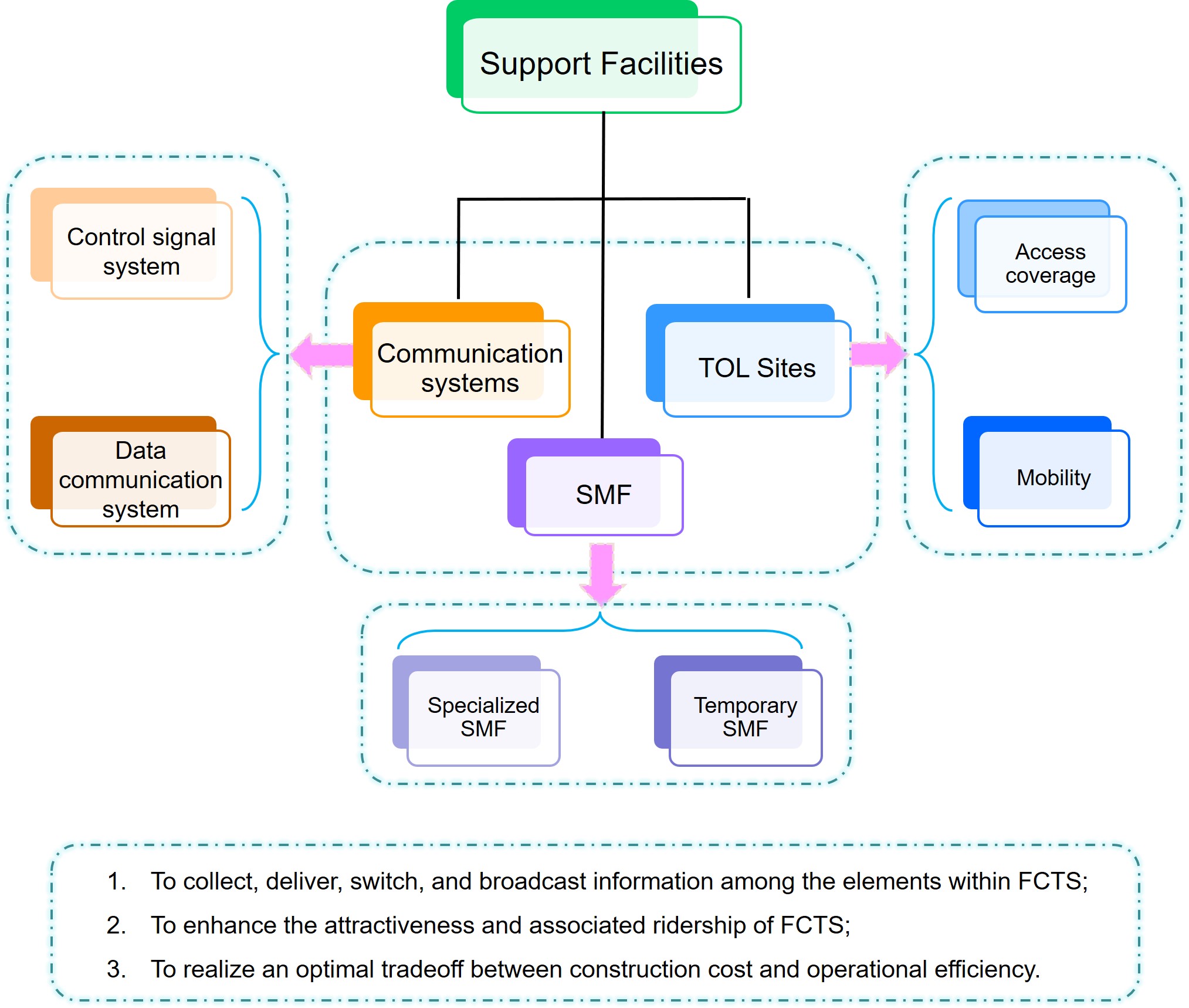}
\caption{Support facilities for FCTS.}
\label{fig_13}
\end{figure*}
Moreover, path/trajectory planning for FCTS for the traditional ground public transportation system must include the passenger experience while traveling in the air that may also be affected by weather conditions along the path/trajectory.

\subsubsection{Fixed Path Planning}
For public transportation in urban areas in which passenger flows are relatively stable, it is commercially viable to provide fixed paths for flying cars, similar to the ground-based public bus transportation system. There are some common design methods and key problems that arise during fixed path planing similar to traditional bus transportation systems (such as traffic load), but some unique features of FCTS should be carefully treated. For example, when choosing TOL sites for flying cars, safety considerations of TOL procedures, noise pollution generated during the TOL process, and forbidden spaces outlined by governments must be taken into account. Moreover, fixed paths should be dynamically maintained and updated according to city developments and the regulations outlined by governments and local authorities.

\subsubsection{Adaptive Path Planning}
Besides public transportation discussed above, there are also some other application scenarios for flying cars in which transportation requirements are dynamic, such as short-distance travel and sightseeing, and customized personalized services. Operators of flying cars should be able to quickly offer the customers convenient and efficient path planning services according to their requirements.

\subsection{Support Facilities}
Similar to traditional ground traffic systems, some support facilities are essential to set up and maintain efficient services of flying cars. Generally, in terms of these different functions, the support facilities of FCTS can be divided into the following: communication systems, TOL sites, and storage and maintenance facilities (SMF), as presented in Fig. \ref{fig_13}.

\subsubsection{Communication Systems}
Communication systems are crucial for FCTS, which serve as the central nervous system that collects, delivers, switches, and broadcasts information between flying cars, control centers, and information service providers. There are two kinds of communication systems that are needed to supply FCTS and passengers with control information and data communication services, respectively, as discussed below.

a) Control signal system: The control signal system for FCTS ensures the efficient organization of flying transportation, allowing flying cars from all participating companies to provided high-quality service, as well as profiting from the service. Through path planning management, TOL sites can be served by flying cars in the event of accidents or path obstructions, or a path re-planning service can be made available for flying cars facing cancellation/unavailability of the previously defined path. Additionally, flying cars can be sent into service rapidly to avoid overcrowding, or even to tend to breakdowns. The monitoring data collected through the control system can also be used to improve timetables and optimize flying car usage.

The key goals of the control signal system for FCTS can be summarized as i) the efficient and intelligent management of public/personal transportation; ii) the optimization of traffic flows leading to better passenger experiences and reductions in energy consumption; iii) integration with other public transportation systems; iv) the promotion of the utilization of FCTS by making it more attractive.

b) Data communication system: The purpose of the data communication system for FCTS is to provide normal voice and multimedia data communications for passengers, which is an essential part of FCTS to enhance the passenger experience during the transportation process. The key requirement of the data communication system for FCTS is that it be reliable, seamless, robust, and provide high-speed data transmissions.

Considering the fact that flying cars are expected to travel at hundreds of meters above ground, the coverage performance of traditional terrestrial wireless communication systems in the operating space of flying cars may not be as good as that on the ground. Therefore, a new communication network design and plan must be provided. For example, high/low altitude platforms and LEO satellite communication systems for 5G and future generation of wireless communications, which aim to provide seamless connections across the world, can be exploited to expand the data communication system for FCTS.

\subsubsection{TOL Sites}
As FCTS are able to provide direct point-to-point transportation services, the locations and construction of TOL sites should enhance the attractiveness and associated ridership of FCTS by providing high-quality services and experience to guarantee efficient mobility for all riders within the service area, especially in urban areas.

Unlike traditional public transportation systems such as public road traffic systems (bus and train) in cities, the ability of flying cars in FCTS to transport large numbers of people is not as significant as buses or trains. Thus, in the initial expansion, FCTS should offer fast and convenient transportation services to premium passengers who can afford the transportation fees, and then once FCTS is established, its expansion can be promoted in the commercial market. Therefore, in urban areas, the locations of TOL sites should be in business zones (e.g., central business districts), offering connecting points to other transportation systems (e.g., bus stops, subway/train stations, airports, etc.), and some other functional places (e.g., hospitals, libraries, schools, etc.).

Specifically, the problem of determining TOL site placements in special areas, such as business zones, is not trivial because a general balance must be achieved between access coverage and mobility. Hence, access coverage and mobility should be taken into consideration simultaneously and methodologies for the determination and optimization of TOL site placements should be proposed at the network level.

The density and locations of TOL sites should typically determine the access coverage of the FCTS service, similarly to the traditional bus transportation system. Some research has revealed that access/egress distance is significantly negatively correlated with transit patronage. Consequently, the spacing of TOL sites should have an upper threshold so as to guarantee that passengers complete their access/egress trips in a reasonable and acceptable amount of time. Generally, such a maximum spacing threshold is location-specific, depending on many local considerations there are (e.g., service regions, road hierarchies, local demands).

Apart from access coverage, mobility is the other prominent factor that greatly affects the system performance and level of service of FCTS. Efficient mobility is preferred, which is closely related to TOL site locations. Given that fewer TOL sites generate less dwelling time at TOL sites and subsequently contribute to higher utilization of the flying cars and lower construction costs of the whole FCTS, minimizing the number of TOL sites is one of the key principles of TOL site deployments.
\begin{figure*}[!htb]
\centering
\includegraphics[width= 4.7in,angle=0]{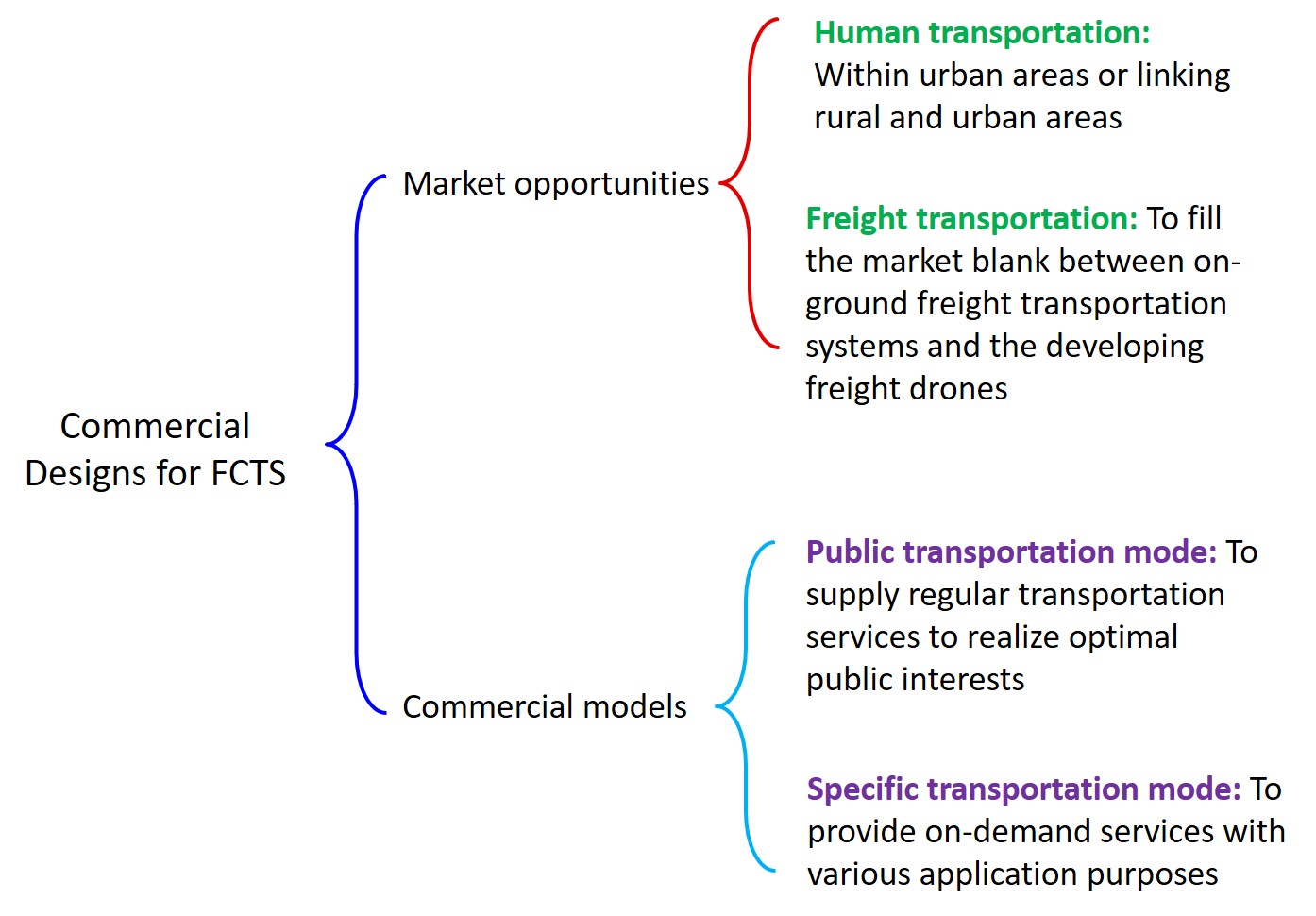}
\caption{Commercial designs for FCTS.}
\label{fig_14}
\end{figure*}
Therefore, these two mutually exclusive factors must be optimally traded-off across the whole FCTS level when considering suitable TOL sites along with the location specifics in a given areas.

\subsubsection{SMF}
Although no roads or tracks are required for FCTS, some necessary supporting facilities are needed to run FCTS, such as SMF \cite{BMFW}, which we discuss below.

There are generally two kinds of SMF related to the unique properties of FCTS and the function of the facilities: 1) Specialized SMF and 2) Temporary SMF. Specialized SMF is tailored to provide daily storage and maintenance for flying cars, and should be large enough to accommodate all flying cars in a service area. Temporary SMF is designed and integrated with normal TOL sites to supply temporary storage and maintenance services for flying cars that are temporarily out of order.

Some important rules for the design and construction of specialized SMF must be addressed. First, SMF should be free of environmental contamination (including noise/light/heavy metals pollution).  Second, SMF should be sufficiently large to accommodate all present requirements and any potential future expansion. Third, SMF should offer comprehensive functions, such as general maintenance and storage, fare service, fueling, exterior washing, and interior cleaning. Lastly, additional space should be provided for staff offices, training/diner/locker rooms, associated toilet facilities, and on-site parking for employees/visitors.

Observing these requirements, SMF can be constructed in suburban areas far from densely populated communities so as to adapt to any constraint factors, including availability and affordability of suitable land and air spaces, real estate, redevelopment interests, and zoning regulations. Therefore, the optimization of the location of SMF should comprehensively address the relationship between construction costs and operational efficiency.

For temporary SMF, the design and construction are not as complicated as those of specialized SMF because temporary storage and maintenance functions can be easily integrated into normal TOL sites. However, temporary SMF unavoidably increase the construction and operation costs of TOL sites, especially in urban areas, leading to a compromise between commercially operation costs and supporting capacity.

\subsection{Commercial Designs}
Because FCTS is a novel transportation method, commercial operation modes must be proposed and designed to outline the commercial viability of FCTS in the real market. In the following section, we will describe the commercial designs of FCTS considering two aspects: Market opportunities and commercial models. The relationship between the main contents is depicted in Fig. \ref{fig_14}.
\begin{figure*}[!htb]
\centering
\includegraphics[width= 7in,angle=0]{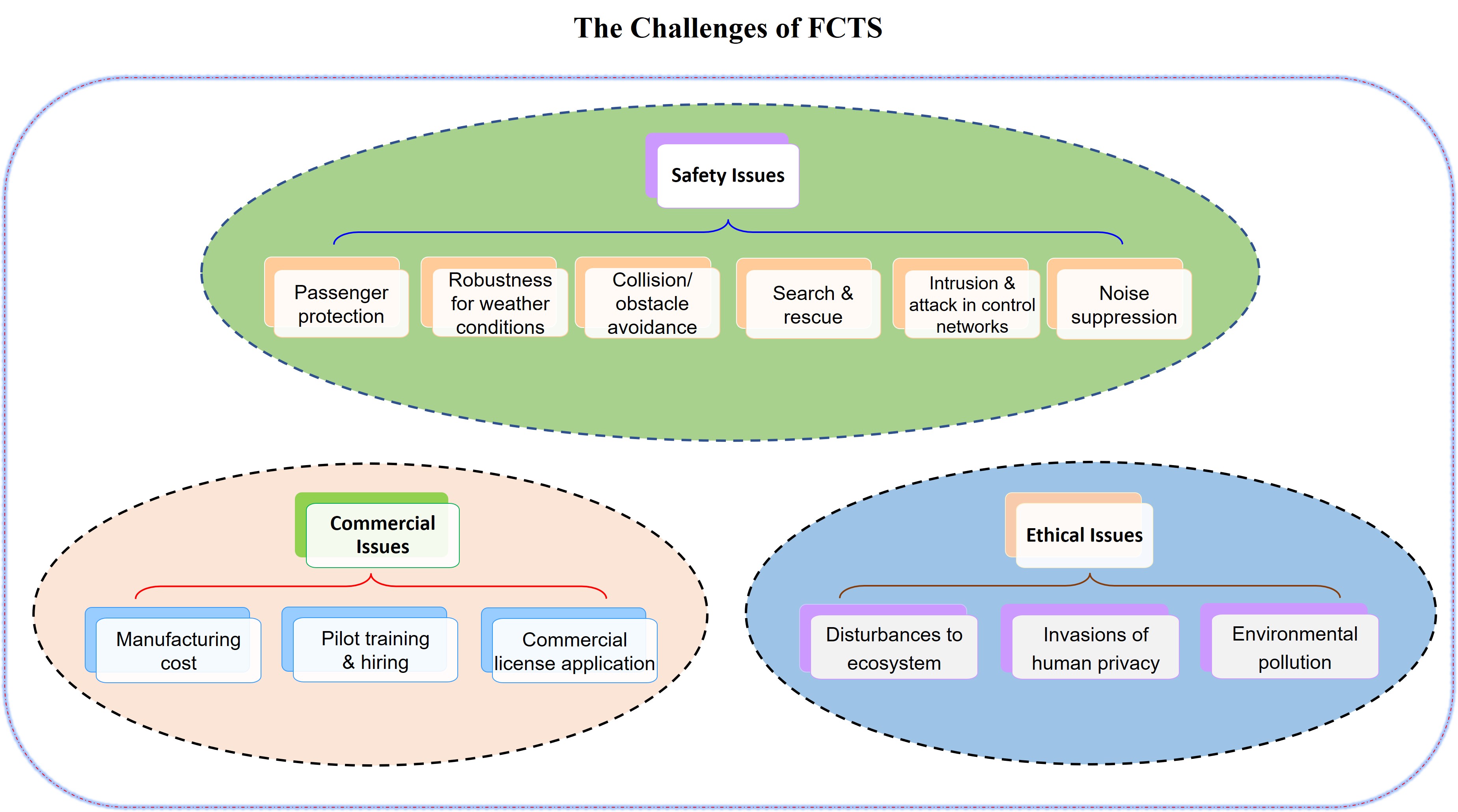}
\caption{The Challenges of FCTS.}
\label{fig_15}
\end{figure*}

\subsubsection{Market Opportunities}
The primary market opportunities of FCTS are related to its ability to relieve the pressure caused by increasing congestion and overcrowding in cities that has been brought about by rapid urban development and the increase in urban populations. Specifically, FCTS is capable of satisfying the main requirements of urban mobility and transportation in cities, which can be loosely categorized into human (public and individual) and freight transportation, while taking the pressure off roads and other infrastructure and reducing transportation costs.

a) Human transportation:
Human transportation is widely regarded as the principal function of FCTS as a means to rapidly transport people in congested cities.  As outlined in previous sections, FCTS can ensure a faster, cheaper, cleaner, and safer transport solution for both public and individual transportation activities, compared to traditional terrestrial transportation systems. FCTS can provide this by combining many technologies, such as ultra-efficient batteries, autonomous driving systems, communication technologies, and advanced manufacturing processes.

b) Freight transportation: Similar to human transportation, FCTS can also offer door-to-door freight transport services, although carrying heavy loads similar to those currently carried by trucks or trains at ground level is not feasible. For example, certain goods that are high priority yet lightweight and of limited size can be rapidly delivered by FCTS. Moreover, the loads that flying cars can carry is much larger than those of drones, for example, which have been used for faster delivery by JD.com \cite{JD} and Amazon \cite{amazon}. Freight flying cars can potentially fulfil the market gap between the traditional on-ground freight transportation systems and the limited freight delivery offered by drones, by simultaneously increasing both the load and delivery range.

\subsubsection{Commercial Models}
In this section, we introduce the commercial models of FCTS and we outline how FCTS can be implemented in the real market. In general, there are two potential types of commercial models for FCTS according to commercial operation behaviors, which we describe below.

a) Public transportation mode: Under this mode, to alleviate traffic congestion issues in cities, FCTS can be used to replace traditional transportation systems, such as public bus/train/subway systems, to transport passengers or goods along fixed routes according to predesigned timetables. FCTS can form part of the public transportation system in a city or urban area, where special operation grants from local governments can be requested. FCTS should be jointly designed and optimized to cooperate with other public transportation systems to maximize optimal public utility. Thus, the construction and operation of FCTS under this mode is a complicated and challenging task, influenced by various factors including local political agendas, real estate and redevelopment interests, city regulation, development and planning, availability of suitable land, conflicts with traditional transportation systems, etc.

b) Specific transportation mode: By benefiting from its uniquely tailorable characteristics, FCTS can be designed for specific application scenarios such as tourism, patrol, monitoring, inspection, anomaly detection/prevention, search and rescue purposes, sightseeing in scenic spots, mine/dam inspection, traffic/environmental monitoring, border/harbor patrol, police surveillance, aerial photography, imaging, and mapping, etc. The design and operation of FCTS under these on-demand application modes are more flexible and convenient, compared with those of FCTS for public transportation applications.

\section{The Challenges of FCTS}
Although several flying car prototypes have already been proposed, manufactured, and tested \cite{palv}-\cite{Audi}, achieving the large-scale practical commercial application of FCTS is still some distance off. Considerable challenges remain, not only technical aspects but also specific commercial and ethical issues \cite{IEEE_EM}, which we illustrate and discuss below. The architecture of the main contents in this section is presented in Fig. \ref{fig_15}.

\subsection{Safety Issues}
The most important challenges faced by the commercial operation of FCTS are safety issues, which arise from various facets such as flying car safety (including passenger protection related to weather conditions), system-level safety (including collision/obstacle avoidance, search and rescue, intrusion and attack in control networks), and environmental safety.

\subsubsection{Passenger Protection}
Unlike traditional terrestrial transportation systems, flying cars normally operate at heights tens to hundreds of meters above ground. Thus, considering the added threat of falling from such heights, potential disasters must be avoided and passengers must be protected from accidents. Flying car malfunctions (such as short circuits or fires) that occur over a busy city, causing the vehicle to fall from above and potentially impacting people below, is a very real threat. Thus, specific survival equipment and emergency mechanisms for FCTS are required. Flying cars should be developed and equipped with specific safety procedures to ensure the safety of passengers, as well as to guarantee the safety of the people and properties below at ground level. These necessary safety requirements for FCTS unavoidably pose numerous engineering challenges for the designers and manufacturers of flying cars.

\subsubsection{Robustness for Weather Conditions}
Because flying cars work in the NGS at various heights, they are inevitably more sensitive to adverse weather conditions in their operating space. The robustness of flying cars to withstand some extreme weather conditions, such as thunderstorms and heavy wind and rain, for example, must be taken into consideration during the design stage. Weather forecasting is also an essential tenet that should be accurately reported to the operator/pilot of flying cars.

\subsubsection{Collision/Obstacle Avoidance}
Because other flying vehicles (such as UAV, helicopters, etc.) share air space with flying cars, collisions may occur between flying cars and other flying vehicles. Therefore, as for traditional aircraft (helicopters and airplanes), flying cars should be equipped with devices that provide early warning and the detection of obstacles in the flight path. On the other hand, an auxiliary ground surveillance network should be implemented to monitor the main operating space of FCTS to provide on-time traffic control and routine schedule services, especially in urban areas.

\subsubsection{Search \& Rescue}
When emergencies such as crashes occur, FCTS should be able to quickly respond by implementing an emergency treatment plan. Specifically, search and rescue operations can be carried out to help injured passengers and the crashed vehicle once the alarm has been raised. Suitable emergency response resources must be optimally prepared, allocated, and distributed in advance in the serving areas of FCTS. The related aspects of search and rescue should form essential components of the system-level design and construction of FCTS.

\subsubsection{Intrusion \& Attack in Control Networks}
In this era of electronic information technology, isolating FCTS from information networks is impossible, and thus the control network of FCTS will be vulnerable to intrusions and attacks. Some protection schemes similar to those adopted in other transportation systems can also be implemented in FCTS to safeguard flying cars and avoid them being hijacked and manipulated.

\subsubsection{Noise Suppression}
Flying cars are expected to operate in urban environments that will inevitably have regulations controlling levels of noise pollution. Designing flying cars to be exceptionally quiet is difficult, especially when they are used for large-scale commercial operations that may be performing hundreds of TOL every hour. Some methods have been proposed to suppress noise, not only from the flying car but also form the CTS. For instance, low noise propellers/electric motors can be optimized for flying cars, and the noise index can be considered in the deployment and design of TOL sites.

\subsection{Commercial Issues}
In this subsection, we discuss the commercial issues encountered by FCTS in large-scale commercial applications.

\subsubsection{Manufacturing Cost}
The manufacturing cost of flying cars is the principal practical challenge that hinders FCTS from being widely commercialized. To achieve the large-scale commercial operation of FCTS, hundreds if not thousands of sustainable and cost-efficient flying cares are needed to satisfy the public transportation requirements in cities to ensure commercial profits. Fortunately, modern manufacturing techniques in the automotive industry can serve as a meaningful reference to significantly reduce the costs of making even complex flying cars by exploiting the integration of new high-tech composites, battery technology, and alloys. Similar to the automotive industry, making flying cars that are affordable for the general public will definitely accelerate the commercialization process of FCTS in both the private and public transportation markets.

\subsubsection{Pilot Training \& Hiring}
Although autonomous driving techniques have been developed for flying cars in self-piloting models, FCTS cannot be completely deprived of pilots, at least in the near future, even when the costs of hiring and training pilots, the salary of whom is typically higher than that of a taxi driver, can be saved. The role of human pilots cannot be replaced for safety reasons in some emergency circumstances. As with self-driving cars, public perception is not yet strong enough as the public are not totally confident with self-piloting techniques. Thus, the full realization of flying cars with entirely self-piloting driving will be a long process, although it will happen eventually. Pilot training and hiring will eventually be phased out, and thus will lead to lower operating cost in FCTS.

\subsubsection{Commercial Licence Application}
Commercial licenses will undoubtedly be necessary for the commercialization of FCTS. Licenses will be issued not only from the economic authorities of the central government but also from the civil aviation regulation department, such as commercial space transportation licenses from Federal Aviation Administration of the United States and European Union Aviation Safety Agency. Furthermore, pilot certificates must be granted to approve a pilot's competence in handling flying cars under all potential flying scenarios, as is the case for traditional aircraft pilots. Therefore, significant effort must be devoted to issuing licences to achieve large-scale commercialization of FCTS, if FCTS is to compete with traditional transportation systems.

\subsection{Ethical Issues}
Because FCTS operate in the NGS, there are some potential ethical issues surrounding the TOL and flying behaviors of flying cars, such as disturbances to ecosystems, invasions of privacy, and pollution to the environment.

\subsubsection{Disturbances to Ecosystem}
Because the height of the operation space for FCTS normally ranges from tens to hundreds of meters height above ground, which is the main living space birds, irresponsible and wanton operation of flying cars will inevitably negatively impact birds by disrupting their nests, provoking attacks, scattering eggs, interrupting feeding, and midair collisions. Furthermore, the noise generated by frequent TOL and flying will also impact ground animals. There are several ways that the negative effects of FCTS can be reduced, such as by monitoring sensitive areas, poaching records, habitat preservation, etc. An environmental impact assessment should be considered during the design and updating process of flying routines, so as to minimize any potential damage to ecosystems and ensure FCTS are eco-friendly.

\subsubsection{Invasions of Human Privacy}
The invasion of human privacy is another problem that might arise from the widespread use of flying cars.  Flying cars that operate over urban or residential areas might leave residents open to being exposed, observed, and supervised in the privacy of their own homes and activities. Flight bans should be enforced to protect sensitive areas, such as the airspace above private properties or prohibited zones.

\subsubsection{Environmental Pollution}
Inadequately controlled noise pollution is the primary pollution produced by FCTS, which can represent a danger to the health and welfare of the population, particularly in urban areas. FCTS should thus obey the noise regulations proposed by local governments, e.g., the Noise Control Act from the environmental protection agency of the United States \cite{AirAct}. Pollution is also introduced by FCTS in the form of the visual pollution of numerous flying cars over the urban landscape. To avoid this, flying routes must be carefully designed and planned.

Furthermore, as discussed in Section III, electricity is the main power source for flying cars, as supplied by batteries that are charged from the grid. Thus, there are two indirect pollution sources resulting from FCTS:

1) Environmental pollution caused by batteries: battery production causes more environmental damage besides carbon emissions alone, and used batteries contain numerous toxic heavy metals. New technologies for green battery production and advanced recycling techniques for waste batteries are required.

2) Emissions during electricity generation: although renewable energy (such as solar, wind, geothermal and modern biofuel production) is the fastest-growing energy source , 76\% of all electricity on the grid across the world and 90\% in the United States in 2018 was generated by traditional fossil fuels including petroleum, natural gas, or coal \cite{c2es}. Therefore, the preferable energy source for FCTS of power batteries still indirectly consumes considerable hydrocarbon fuels. Thus, green power generation on the country/world level is needed to reduce emissions.

\section{Conclusion}
In this paper, we have presented a comprehensive study on FCTS in five focused subject areas. First, we reviewed the  history of the development of flying cars in terms of both timeline and the technique categories. Second, we discussed and compared the main existing designs of flying cars in four aspects: TOL modes, pilot modes, operation modes, and power types. We also introduced the designs of FCTS by considering path and trajectory planning, supporting facilities, and commercial designs. Moreover, we provided the challenges and future developing directions of FCTS in detail.

Based on the discussions and comparisons among the main existing designs of flying cars, some general guidelines for the future design and manufacture of flying cars are:

\begin{itemize}
\item Emergency mechanisms must be provided and safety equipment must be installed on flying cars, as the safety of passengers is the absolute priority for the commercialization of flying cars;

\item Flying cars should be sufficiently robust to operate in all weather conditions;

\item VTOL and electric power are necessary for flying cars operating in urban areas, to adapt to space-constrained urban areas and improve environmental friendliness;

\item Flying cars with HTOL is a good choice for some special application scenarios where enough space is available to build runways;

\item Hydrocarbon fuel and hybrid power are suitable for flying cars operating over long distances;

\item Autonomous pilot is the best choice for flying cars operating along fixed lines and time schedules, similarly to public bus and train services, due to the saved operational costs without using a human pilot;

\item Similar to airplanes, the hybrid pilot mode is a better choice for normal flying cars because it doesn't fully rely on human pilots except in emergencies.
\end{itemize}

Furthermore, according to the presented discussions on the designs and challenges of FCTS, we highlight some useful principles for the construction and commercialization of FCTS below:

\begin{itemize}
\item A collision/obstacle avoidance mechanism must be designed for FCTS, while a search \& rescue program must be predefined before FCTS is commercialized;

\item Green manufacture and operation should be encouraged and advocated in FCTS to avoid any unnecessary environmental pollution;

\item New and accurate position and navigation algorithms are required for FCTS, which operates in 3D space, unlike traditional grounded transportation systems that working in 2D space;

\item Path/trajectory planning should be conducted and optimized for specific operation scenarios, while considering some constraints, such as weather conditions and conservation spaces;

\item The communication system for FCTS should be specialized to provide maximum coverage performance in 3D NGS, while taking the information security problem into consideration during the design of the control signal system for FCTS;

\item The locations of TOl sites and SMF should be jointly chosen, emerged, and optimized in specific serving areas, where the functions of TOL and SMF can be launched in the same physical space;

\item The main market opportunity for FCTS is public transportation, which can significantly reduce traffic congestion in urban areas; there are also some specific application scenarios for FCTS that require specialized on-demand services;

\item Manufacturing and operation costs should be controlled and saved to ensure the survival of FCTS in the market when competing with traditional transportation systems;

\item More effort must be made to gather support from investors but from the government to achieve the financial and policy support, respectively, and to guarantee the successful commercialization of FCTS in the future.

\end{itemize}

\end{document}